\documentclass[aps,prl,twocolumn,amsmath,amssymb,floatfix]{revtex4-2}

\usepackage{graphicx}
\usepackage{bm}
\usepackage{hyperref}
\hypersetup{colorlinks=true,linkcolor=blue,citecolor=blue,urlcolor=blue}
\usepackage{orcidlink}

\newcommand{\Mp}{M_P}
\newcommand{\lcd}{\bar{\nabla}}
\newcommand{\lcR}{\bar{R}}
\newcommand{\Tt}{\tilde{T}}
\newcommand{\Qt}{\tilde{Q}}
\newcommand{\Ut}{\tilde{U}}
\newcommand{\Qh}{\hat{Q}}
\newcommand{\Rs}{\mathring{R}}   % invariant curvature; change here only (e.g. \hat{R})

\begin{document}

\title{The metric-affine $\Rs^2$ scalaron: parameter-free inflation}

\author{Christian Dioguardi\,\orcidlink{0000-0002-6133-0383}}
\email{christian.dioguardi@kbfi.ee} 
\affiliation{Laboratory of High Energy and Computational Physics, National Institute of Chemical Physics and Biophysics, R\"avala pst.~10, Tallinn, 10143, Estonia}

\begin{abstract}
We present a model of slow-roll inflation in metric-affine gravity, where the connection is independent of the metric. We require that the Einstein--Hilbert term is the only operator that breaks three out of four shift-symmetries of the connection, and impose them on the metric-affine Starobinsky term $\alpha R(\Gamma)^2$. This replaces it by $\alpha\Rs^2$, the square of the invariant part of the scalar curvature. The theory propagates only the graviton and one healthy scalar, the metric-affine $\Rs^2$ scalaron. Its potential approaches a plateau polynomially, $U\simeq U_0(1-\mu^2/\chi^2)$, with $\mu=\sqrt{6}\,M_P$ fixed by the curvature of the connection. Once $\alpha\simeq2\times10^{8}$ is set by the scalar amplitude, the model has no free parameters. It predicts $n_s\simeq1-3/(2N_*)$ and $r\simeq2\sqrt3\,N_*^{-3/2}$ at leading order in $N_*$, and $n_s=0.968$--$0.974$ and $r=0.011$--$0.008$ with exact slow-roll for $N_*=50$--$60$ $e$-folds. These values agree with current data, and $r$ is within reach of the next generation of $B$-mode measurements.
\end{abstract}

\maketitle

%======================================================================
\section{Introduction} Inflation, a phase of accelerated expansion in the early Universe, explains the flatness and homogeneity of the observed Universe and generates the primordial perturbations from quantum fluctuations~\cite{Starobinsky:1980te,Guth:1980zm,Linde:1981mu,Albrecht:1982wi,Mukhanov:1981xt,Starobinsky:1982ee,Guth:1982ec,Bardeen:1983qw}. In single-field slow-roll models these perturbations are described by three main observables: the amplitude $A_s$ and tilt $n_s$ of the scalar spectrum, and the tensor-to-scalar ratio $r$.

Measurements of the Cosmic Microwave Background (CMB) now determine $n_s$ to a few per mille. A joint analysis of Planck~\cite{Planck:2018jri}, ACT~\cite{ACT:2025tim}, SPT~\cite{SPT-3G:2025bzu} and BICEP/Keck~\cite{BICEP:2021xfz} finds $n_s=0.9682\pm0.0032$ and $r<0.034$ (95\% CL) from the CMB alone, and $n_s=0.9728\pm0.0029$ and $r<0.035$ when DESI BAO~\cite{DESI:2025zgx} is included~\cite{Balkenhol:2025wms}.

The bound on $r$ comes from BICEP/Keck $B$-mode data, $r<0.036$~\cite{BICEP:2021xfz}, and is expected to improve soon. The next BICEP/Keck analysis, with data through 2024, is projected to reach $\sigma(r)\simeq0.003$~\cite{Pryke:2025talk}. The Simons Observatory (SO) ~\cite{SimonsObservatory:2018koc}, and BICEP Array \cite{BICEPArray:2026status} projects forecast $\sigma(r)\simeq 10^{-3}$ by 2035.
A model that predicts $r\sim10^{-2}$ can already be tested by the next BICEP/Keck analysis.

Starobinsky inflation~\cite{Starobinsky:1980te} is currently the benchmark model of inflation. The action
\begin{equation}
 S = \int d^4x\sqrt{-g}\,\big(\tfrac12\Mp^2R+\alpha R^2\big)\,,
\end{equation}
is equivalent to General Relativity (GR) plus one scalar, the scalaron~\cite{Whitt:1984pd}, with potential $U=U_0\big(1-e^{-\sqrt{2/3}\,\chi/\Mp}\big)^2$ and $U_0=\Mp^4/(16\alpha)$. Once $\alpha\approx5\times10^8$ is fixed by $A_s$, the model has no free parameters: $n_s\simeq1-2/N_*$ and $r\simeq12/N_*^2$, where $N_*$ is the number of $e$-folds between the horizon exit of the pivot scale and the end of inflation. Higgs inflation gives, at leading order, the same predictions~\cite{Bezrukov:2007ep}. In the Starobinsky action only $R^2$ is kept, although the Weyl term $C_{\mu\nu\rho\sigma}C^{\mu\nu\rho\sigma}$ is allowed at the same order. The Weyl term propagates a massive spin-2 ghost~\cite{Stelle:1976gc,Stelle:1977ry}; with a coefficient of natural size its mass is of order $\Mp$, far above the inflationary scale. For $N_*=50$--$60$, Starobinsky inflation gives $n_s=0.962$--$0.968$ and $r=0.0042$--$0.0030$. This $n_s$ lies $1.7\sigma$ to $3.9\sigma$ below the CMB+BAO value, and $r$ is beyond the reach of the next BICEP/Keck release.

In this Letter we construct the analogue of Starobinsky inflation in metric-affine gravity, where the connection is independent of the metric~\cite{Hehl:1994ue,Sotiriou:2006qn}. Two obstacles appear. In the Palatini formulation, $R(\Gamma)^2$ does not propagate a scalar~\cite{Enckell:2018hmo,Antoniadis:2018ywb}. On the other hand, generic unconstrained curvature-squared terms of an independent connection propagate ghosts~\cite{BeltranJimenez:2019acz,BeltranJimenez:2020sqf,Percacci:2019hxn}. Specific terms avoid this, such as the square of the Holst invariant, which propagates a healthy pseudoscalar~\cite{Salvio:2022suk}; for this and other inflationary models with a dynamical connection see Refs.~\cite{Salvio:2022suk,Gialamas:2022xtt,Racioppi:2024pno,Racioppi:2025igu}. Here, we overcome both obstacles with a symmetry principle: four local shifts of the connection (see, for example, Refs.~\cite{Iosifidis:2019fsh,Sauro:2022hoh,Ferrara:2026bzy}), three are broken only by the Einstein--Hilbert term. Imposed on the metric-affine Starobinsky term $R(\Gamma)^2$, the shifts turn it into $\Rs^2$, the square of the invariant part of the scalar curvature, which propagates exactly one healthy scalar, the metric-affine scalaron.

The potential generated in this way has the same plateau as in Starobinsky inflation but approaches it polynomially, $U\simeq U_0(1-6\Mp^2/\chi^2)$. Potentials of this form appear in polynomial $\alpha$-attractors and brane inflation, with the coefficient of $1/\chi^2$ a free parameter~\cite{Kallosh:2022feu,Martin:2013tda,Dioguardi:2023jwa,Dioguardi:2025vci}. In the language of pole attractors~\cite{Galante:2014ifa}, the scalaron kinetic term has a pole of order three, which gives $n_s\simeq1-3/(2N_*)$. Here the curvature of the connection fixes its residue, and with it the coefficient of $1/\chi^2$. The model predicts $n_s\simeq1-3/(2N_*)$ and $r\simeq2\sqrt3\,N_*^{-3/2}$ at leading order, with no free parameters.

%======================================================================
\section{Metric-affine setup} We use the signature $(-,+,+,+)$ and $\epsilon_{0123}=\sqrt{-g}$, and denote with $\Mp$ the reduced Planck mass. The connection is decomposed as
\begin{equation}
\Gamma^\lambda{}_{\mu\nu}=\bar\Gamma^\lambda{}_{\mu\nu}+N^\lambda{}_{\mu\nu}\,,
\end{equation}
with covariant derivative $\nabla_\mu V^\lambda=\partial_\mu V^\lambda+\Gamma^\lambda{}_{\mu\nu}V^\nu$ and curvature 
\begin{equation}
R^\rho{}_{\sigma\mu\nu}=\partial_\mu\Gamma^\rho{}_{\nu\sigma}-\partial_\nu\Gamma^\rho{}_{\mu\sigma}+\Gamma^\rho{}_{\mu\lambda}\Gamma^\lambda{}_{\nu\sigma}-\Gamma^\rho{}_{\nu\lambda}\Gamma^\lambda{}_{\mu\sigma}\,.   
\end{equation}
Here $\bar\Gamma$ is the Levi-Civita connection and $N$ is the distortion tensor, which is fixed by the torsion $T^\lambda{}_{\mu\nu}=\Gamma^\lambda{}_{\mu\nu}-\Gamma^\lambda{}_{\nu\mu}$ and the non-metricity $Q_{\lambda\mu\nu}=\nabla_\lambda g_{\mu\nu}$, see Eq. \eqref{eq:distortion} of the End Matter. The 64 components of $N$ split into four independent vectors,
\begin{align}
  T_\mu&=T^\lambda{}_{\lambda\mu}, & Q_\mu&=Q_{\mu\lambda}{}^{\lambda},\nonumber\\
  \Qt_\mu&=Q^\lambda{}_{\lambda\mu}, & \Tt^\mu&=\epsilon^{\mu\nu\rho\sigma}T_{\nu\rho\sigma},
\end{align}
where $\Tt$ is an axial vector, and 48 tensor components: the traceless parts $\hat T_{\lambda\mu\nu}$ of torsion (16 components) and $\Qh_{\lambda\mu\nu}$ of non-metricity (32 components), defined in the End Matter, Eqs.\eqref{eq:hats}-\eqref{eq:hats_2}.

We consider four local transformations of the connection, built from the structures $\delta^\lambda_\nu$, $\delta^\lambda_\mu$, $g_{\mu\nu}$ and $\epsilon^\lambda{}_{\mu\nu\rho}$:
\begin{equation}
\begin{aligned}
  &\text{projective:}   && \delta\Gamma^\lambda{}_{\mu\nu}=\delta^\lambda_\nu\,\xi_\mu ,\\
  &\text{trace:}        && \delta\Gamma^\lambda{}_{\mu\nu}=\delta^\lambda_\mu v_\nu-2g_{\mu\nu}v^\lambda ,\\
  &\text{metric-trace:} && \delta\Gamma^\lambda{}_{\mu\nu}=g_{\mu\nu}X^\lambda,\quad X^\lambda=\lcd_\rho A^{\rho\lambda} ,\\
  &\text{axial:}        && \delta\Gamma^\lambda{}_{\mu\nu}=\epsilon^\lambda{}_{\mu\nu\rho}Y^\rho .
\end{aligned}
\label{eq:shifts}
\end{equation}
With $A^{\rho\lambda}=-A^{\lambda\rho}$, the metric-trace parameter is transverse, $\lcd_\lambda X^\lambda=0$; all other parameters are unconstrained.
The projective transformation is the standard symmetry of the Einstein--Hilbert term~\cite{Hehl:1994ue,Sotiriou:2006qn,Afonso:2017bxr}, while the others have been studied in~\cite{Iosifidis:2019fsh,Sauro:2022hoh,Ferrara:2026bzy}.
All four leave $\hat T$ and $\Qh$ invariant. The projective transformation also leaves invariant the three vectors
\begin{align}
  \Sigma_\mu&=T_\mu-\tfrac5{12}Q_\mu+\tfrac16\Qt_\mu,\\
  \Ut_\mu&=-\tfrac32T_\mu+\tfrac38Q_\mu+\tfrac34\Qt_\mu,
\end{align}
and $\Tt_\mu$. Each of the other three transformations moves exactly one of them:
\begin{equation}
  \delta\Sigma_\mu=3v_\mu,\qquad \delta\Ut_\mu=-\tfrac92X_\mu,\qquad \delta\Tt_\mu=12Y_\mu .
\end{equation}

The scalar curvature $R(\Gamma)=g^{\sigma\nu}R^\rho{}_{\sigma\rho\nu}$ decomposes as
\begin{multline}
  R(\Gamma)=\lcR+\hat{R}-\Sigma_\mu\Sigma^\mu+\tfrac4{27}\Ut_\mu\Ut^\mu+\tfrac1{24}\Tt_\mu\Tt^\mu\\
  -\lcd_\mu\big(3\Sigma^\mu+\tfrac23\Ut^\mu\big),
  \label{eq:Rdec}
\end{multline}
where $\lcR$ is the Levi-Civita scalar curvature and $\hat{R}$ is a quadratic form in $\hat T$ and $\Qh$ [see Eq.~\eqref{eq:quadratic} in the End Matter]. $\bar R$ is trivially invariant. $\hat{R}$ invariance follows from invariance of $\hat T,\hat Q$ under all four transformations. The three vector bilinears are not invariant, and neither is $\lcd_\mu\Sigma^\mu$. The divergence $\lcd_\mu\Ut^\mu$ is invariant because $X^\mu$ is transverse.

Equation~\eqref{eq:Rdec} contains no products of vectors moved by different shifts: the Einstein--Hilbert term breaks the shifts independently. Among the shifts built from $\delta^\lambda_\mu$ and $g_{\mu\nu}$, this condition fixes the trace shift uniquely, given the metric-trace shift. With $\delta\Gamma^\lambda{}_{\mu\nu}=\delta^\lambda_\mu v_\nu$ instead, the theory below would contain a ghost.

Let $\Pi$ delete from a local scalar every monomial in $\Sigma$, $\Ut$, $\Tt$, $\hat T$, $\Qh$, $g_{\mu\nu}$ and their $\lcd$-derivatives, decomposed into irreducible parts, that is not invariant under Eq.~\eqref{eq:shifts}. A product of monomials is invariant only if each factor is, so $\Pi$ is multiplicative, $\Pi[AB]=\Pi[A]\,\Pi[B]$. Applied to the scalar curvature, it defines the invariant curvature
\begin{equation}
  \Rs\equiv\Pi[R(\Gamma)]=\lcR+\hat{R}-\tfrac23\lcd_\mu\Ut^\mu .
  \label{eq:Rstar}
\end{equation}
Any function of $\Rs$ is invariant under the four transformations exactly, not only up to boundary terms, and $\Pi[R(\Gamma)^n]=\Rs^n$.
%======================================================================
\section{The action}
The guiding principle is that gravity is General Relativity at leading order and that all corrections respect the shift symmetries, so that the Einstein--Hilbert term is the only operator that breaks them. We start from the metric-affine analogue of Starobinsky's action, $\tfrac12\Mp^2R(\Gamma)+\alpha R(\Gamma)^2$, and impose the shifts on the curvature-squared term, which replaces $R(\Gamma)^2$ by its invariant part, $\Pi[R(\Gamma)^2]=\Rs^2$:
\begin{equation}
\begin{aligned}
  S&=\int d^4x\sqrt{-g}\,\Big[\tfrac12\Mp^2R(\Gamma)+\alpha\,\Pi[R(\Gamma)^2]\Big]\\
   &=\int d^4x\sqrt{-g}\,\Big[\tfrac12\Mp^2R(\Gamma)+\alpha\Rs^2\Big].
\end{aligned}
  \label{eq:action}
\end{equation}

Note that the transversality of $X^\mu$ is the key ingredient. If the metric-trace shift is not restricted to transverse $X$, $\lcd\!\cdot\!\Ut$ is not invariant and $\Pi[R(\Gamma)^2]=(\lcR+\hat R)^2$.  In that case the field equations set $\Sigma=\Ut=\Tt=\hat T=\Qh=0$, so the connection is Levi-Civita up to a projective transformation and the same construction gives Starobinsky inflation.

For $\alpha\neq0$ the action~\eqref{eq:action} is equivalent to~\cite{Whitt:1984pd}
\begin{equation}
  S=\int d^4x\sqrt{-g}\,\Big[\tfrac12\Mp^2R(\Gamma)+\tfrac12\Omega\,\Rs-\frac{\Omega^2}{16\alpha}\Big],
  \label{eq:aux}
\end{equation}
where the equation of motion of the auxiliary field $\Omega$ gives $\Omega=4\alpha\Rs$; we also define $F(\Omega) = M_P^2 + \Omega$.

Three remarks motivate this action.

First, the explicit breaking is what makes the theory healthy. By Eq.~\eqref{eq:Rdec}, $\tfrac12\Mp^2R(\Gamma)$ contains the vector terms $-\tfrac12\Mp^2\Sigma^2+\tfrac2{27}\Mp^2\Ut^2+\tfrac1{48}\Mp^2\Tt^2$. The trace shift forbids invariant operators from sourcing $\Sigma$, the only vector with a negative coefficient. The metric-trace and axial shifts forbid field-dependent corrections to the masses of $\Ut$ and $\Tt$. As shown below, this makes the kinetic term generated by the distortion positive. The breaking cannot be removed: with $\tfrac12\Mp^2\Rs$ instead of $\tfrac12\Mp^2R(\Gamma)$, the vectors have no mass terms, $\Ut$ enters only through $\tfrac13\partial_\mu\Omega\,\Ut^\mu$ and acts as a Lagrange multiplier that freezes $\Omega$, and the scalaron does not propagate.

Second, $R(\Gamma)^2$ itself is excluded. It is equivalent to a scalar coupled to $R(\Gamma)$, a field-dependent breaking of the shifts that sources $\Sigma$ and gives $\Ut$ a field-dependent mass. Integrating out the distortion then gives the Jordan-frame term $+\tfrac{3}{4F(\Omega)}(\partial\Omega)^2$, which cancels exactly the conformal contribution in the Einstein frame. This is the Palatini case, where the scalaron does not propagate~\cite{Enckell:2018hmo,Antoniadis:2018ywb}.

Third, the shifts allow other curvature-squared terms, such as $\lcR^2$, $(\lcd\cdot\Ut)^2$, and squares of the curvature of the connection $\bar\Gamma+N[\hat T,\Qh]$. As in Starobinsky inflation, where $C^2$ is allowed but assumed to have a natural coefficient, we assume that $R(\Gamma)^2$ is the operator with a large coefficient.  The other terms with $\mathcal O(1)$ coefficients generically give modes with masses of order $\Mp$, far above the Hubble rate during inflation. The shifts alone allow $\lcR+\hat R+\kappa\,\lcd\!\cdot\!\Ut$ for any $\kappa$, since $\lcR$ and $\lcd\!\cdot\!\Ut$ are separately invariant. The coefficient $\kappa=-2/3$ chosen here is the coefficient of the only invariant derivative of the connection in $R(\Gamma)$. Other values of $\kappa$ require $\lcR\,\lcd\!\cdot\!\Ut$ and $(\lcd\!\cdot\!\Ut)^2$ with coefficients of order $\alpha$, which are not part of $R(\Gamma)^2$. With natural coefficients these operators shift $\kappa$ by $\mathcal O(1/\alpha)\sim10^{-8}$.\\

%======================================================================
Using Eqs.~\eqref{eq:Rdec}--\eqref{eq:Rstar} and integrating by parts, the Lagrangian becomes
\begin{multline}
  \frac{\mathcal L}{\sqrt{-g}}=\tfrac12F(\Omega)\,(\lcR+\hat{R})-\tfrac12\Mp^2\Sigma_\mu\Sigma^\mu+\tfrac2{27}\Mp^2\Ut_\mu\Ut^\mu\\
  +\tfrac1{48}\Mp^2\Tt_\mu\Tt^\mu+\tfrac13\,\partial_\mu\Omega\,\Ut^\mu-\frac{\Omega^2}{16\alpha}.
  \label{eq:Ldist}
\end{multline}
The distortion enters algebraically, so we can integrate it out. The tensor parts, $\Sigma$ and $\Tt$ have no sources and vanish: $\hat T=\Qh=0$ for $F(\Omega)\neq0$, because $\hat R$ is non-degenerate (see End Matter), and $\Sigma=\Tt=0$. The vector $\Ut$ follows $\Omega$,
\begin{equation}
  \Ut_\mu=-\frac{9}{4\Mp^2}\,\partial_\mu\Omega .
\end{equation}
Substituting back,
\begin{equation}
  \frac{\mathcal L}{\sqrt{-g}}=\tfrac12F(\Omega)\lcR-\frac{3}{8\Mp^2}(\partial\Omega)^2-\frac{\Omega^2}{16\alpha}.
  \label{eq:Jordan}
\end{equation}
The distortion adds the kinetic term in Eq.~\eqref{eq:Jordan}, with the healthy sign. The sign is fixed by the positive coefficient of $\Ut^2$ in Eq.~\eqref{eq:Ldist} and by the absence of a $\Sigma$ source. A source along $\Sigma$, as in the Palatini case, would contribute with the opposite sign.

%======================================================================
The non-minimal coupling to $\lcR$ is removed by the conformal transformation 
\begin{equation}
\tilde g_{\mu\nu}=\frac{F(\Omega)}{\Mp^2}\,g_{\mu\nu}.    
\end{equation} With $\theta\equiv\Omega/\Mp^2$, and dropping the tilde, the action in the Einstein frame is
\begin{multline}
  S=\int d^4x\sqrt{-g}\,\bigg[\frac{\Mp^2}2\lcR-\frac{3\Mp^2}{8}\,\frac{3+\theta}{(1+\theta)^2}(\partial \theta)^2\\
  -U_0\frac{\theta^2}{(1+\theta)^2}\bigg],
  \label{eq:Einstein}
\end{multline}
with $U_0\equiv\Mp^4/(16\alpha)$. Relative to Starobinsky inflation, the distortion multiplies the scalaron kinetic term by $(3+\theta)/2$. The scalar kinetic term is positive for $\theta>-3$; we require $F(\Omega)>0$, i.e.\ $\theta>-1$, for a healthy graviton and a regular conformal transformation, and $\alpha>0$ for $U_0>0$.

The canonical field $\chi$ obeys
\begin{equation}
  \Big(\frac{d\chi}{d\theta}\Big)^2=\frac{3}{4}\,\Mp^2\,\frac{3+\theta}{(1+\theta)^2}.
\end{equation}
With $w\equiv\sqrt{3+\theta}$ and up to an additive constant, the canonical field $\chi$ takes a simple analytical form:
\begin{equation}
  \chi=\sqrt3\,\Mp\Big[w-\sqrt2\,\mathrm{artanh}\frac{\sqrt2}{w}\Big],\qquad
  U=U_0\Big(\frac{w^2-3}{w^2-2}\Big)^2.
  \label{eq:canonical}
\end{equation}
The potential has a minimum at $\theta=0$ ($w=\sqrt{3}$, i.e.\ $\chi\simeq0.19\,\Mp$ with this choice of the constant), where the scalaron mass is $m_\chi^2=\Mp^2/(18\alpha)$. At large $\chi$, $w\simeq\chi/(\sqrt3\Mp)$ and
\begin{equation}
  U(\chi)\simeq U_0\Big(1-\frac{6\Mp^2}{\chi^2}+\dots\Big).
  \label{eq:plateau}
\end{equation}
The Starobinsky potential approaches the same plateau exponentially. Here the approach is polynomial, and its coefficient, $\mu^2=6\Mp^2$, is fixed. In terms of $\rho\equiv1/\theta$, the kinetic function in Eq.~\eqref{eq:Einstein} has a pole of order three at $\rho=0$ with residue $3/4$, in the sense of Ref.~\cite{Galante:2014ifa}. 

%======================================================================
\section{Slow-roll predictions}

We use the standard slow-roll formalism with
\begin{equation}
\epsilon_U\equiv\tfrac12\Mp^2(U_\chi/U)^2\,, \qquad \eta_U\equiv\Mp^2U_{\chi\chi}/U\,,
\end{equation}
in which the observables at horizon exit of the pivot scale $\chi=\chi_*$, are
\begin{align}
n_s&=1-6\epsilon_U(\chi_*)+2\eta_U(\chi_*)\,,\\ r&=16\epsilon_U(\chi_*)\,,\\ A_s&=\frac{U(\chi_*)}{24\pi^2\Mp^4\epsilon_U(\chi_*)}\,,
\end{align}
and the number of $e$-folds between horizon exit and the end of inflation is
\begin{equation}
N_*=\frac{1}{\Mp^2}\int_{\chi_{\rm end}}^{\chi_*} \frac{U}{U_\chi}\,d\chi\,.
\end{equation}
In terms of $\theta$ all of them are elementary functions:
\begin{equation}\label{eq:epseta}
  \epsilon_U=\frac{8}{3\theta^2(\theta+3)},\qquad \eta_U=-\frac{4(3\theta^2+5\theta-6)}{3\theta^2(\theta+3)^2}\,,
\end{equation}
\begin{align}\label{eq:nsr}
 n_s&=1-\frac{8(3\theta_*^2+11\theta_*+12)}{3\theta_*^2(\theta_*+3)^2},\\ \nonumber r&=\frac{128}{3\theta_*^2(\theta_*+3)},\;\; A_s=\frac{\theta_*^4(\theta_*+3)}{1024\pi^2\alpha\,(1+\theta_*)^2}\,.
  \label{eq:As}    
\end{align}

The number of $e$-folds is $N_*=G(\theta_*)-G(\theta_{\rm end})$, with
\begin{equation}
  G(\theta)=\tfrac3{16}\big[\theta^2+4\theta-4\ln(1+\theta)\big].
\end{equation}
Inflation ends at $\epsilon_U=1$, i.e.\ $\theta_{\rm end}\simeq0.834$ and $G(\theta_{\rm end})\simeq0.30$.

For $N_*\gg1$ we can neglect the end of inflation, $G(\theta_{\rm end})\to0$, and invert:
\begin{equation}
\theta_*=4\sqrt{N_*/3}-2+\mathcal O(N_*^{-1/2}\ln N_*).
\end{equation}
To next-to-leading order in $N_*^{-1/2}$,
\begin{align}
  n_s&=1-\frac{3}{2N_*}-\frac{5\sqrt3}{8N_*^{3/2}},\label{eq:nsN}\\
  r&=\frac{2\sqrt3}{N_*^{3/2}}\bigg(1+\frac{3\sqrt3}{4\sqrt{N_*}}\bigg),\label{eq:rN}\\
  A_s&=\frac{N_*^{3/2}}{48\sqrt3\,\pi^2\alpha}\bigg(1-\frac{5\sqrt3}{4\sqrt{N_*}}\bigg).\label{eq:AsN}
\end{align}
The end of inflation and the logarithmic terms enter only at the following order. Eliminating $N_*$ at leading order gives
\begin{equation}
  r\simeq2\sqrt3\,\Big[\tfrac23(1-n_s)\Big]^{3/2},
  \label{eq:relation}
\end{equation}
compared with $r\simeq3(1-n_s)^2$ for Starobinsky inflation. A potential $U_0(1-\mu^2/\chi^2)$ gives $n_s\simeq1-3/(2N_*)$ for any $\mu$, and $r\simeq\sqrt2\,(\mu/\Mp)N_*^{-3/2}$~\cite{Kallosh:2022feu}, so Eq.~\eqref{eq:relation} corresponds to $\mu=\sqrt6\,\Mp$. Evaluated at the exact slow-roll value of $n_s$, Eq.~\eqref{eq:relation} reproduces the exact slow-roll $r$ to $1.5\%$ for $N_*=50$--$60$. Because Eq.~\eqref{eq:relation} does not depend on $N_*$, measuring $n_s$ and $r$ tests $\mu$ without input from reheating.
For $N_*=50$--$60$, the normalization $A_s=2.1\times10^{-9}$ fixes
\begin{equation}
 \alpha=(1.6-2.2)\times10^8,\qquad m_\chi\simeq4\times10^{13}\,\rm{GeV}.   
\end{equation}

\begin{figure}[t]
\centering
\includegraphics[width=\columnwidth]{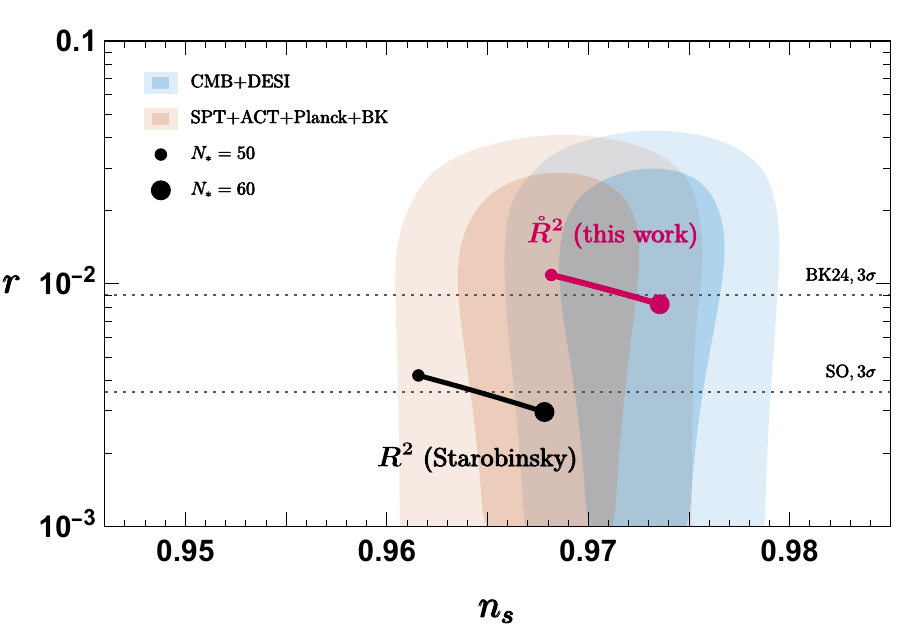}
\caption{Predictions for $N_*=50$--$60$ for the $\Rs^2$ model (magenta) and Starobinsky inflation (black), in exact slow roll. The shaded regions show the 68\% and 95\% CL from Planck, BICEP/Keck, ACT and SPT~\cite{Balkenhol:2025wms} without (orange) and with (blue) DESI BAO~\cite{DESI:2025zgx}. Dotted lines: $r=3\sigma(r)$ for the next BICEP/Keck analysis, $\sigma(r)\simeq0.003$~\cite{Pryke:2025talk} (BK), and for Simons Observatory (SO), $\sigma(r)\simeq1.2\times10^{-3}$~\cite{SimonsObservatory:2025sat}.}
\label{fig:nsr}
\end{figure}

Figure~\ref{fig:nsr} shows the predictions with current constraints and forecasts. For $N_*=50$--$60$, $n_s=0.968$--$0.974$ and $r=0.0109$--$0.0082$. The spectral index lies within $1.7\sigma$ of both the CMB-only and the CMB+BAO values~\cite{Balkenhol:2025wms}, and within $1\sigma$ of the latter for $N_*\gtrsim53$. The tensor-to-scalar ratio is a factor 3--4 below the current bound. At $N_*=55$, $r=0.0094$ would appear at about $3\sigma$ in the next BICEP/Keck analysis and at about $7-8\sigma$ in the SO survey.

%======================================================================
\section{Conclusions} We have shown that metric-affine gravity contains a healthy analogue of Starobinsky inflation, which propagates only the graviton and one scalar. The key ingredient is a symmetry principle: four local shifts of the connection, three broken only by the Einstein--Hilbert term. Imposed on the metric-affine Starobinsky term $R(\Gamma)^2$, this principle selects $\Rs^2$, and the action $\tfrac12\Mp^2R(\Gamma)+\alpha\Rs^2$ propagates one scalar, the metric-affine scalaron. The potential keeps the Starobinsky plateau but approaches it as $U_0(1-\mu^2/\chi^2)$: the curvature of the connection fixes the plateau coefficient to $\mu=\sqrt6\,\Mp$.

With $\alpha$ fixed by $A_s$, the model has no free parameters. It predicts $n_s\simeq1-3/(2N_*)$ and $r\simeq2\sqrt3\,N_*^{-3/2}$ at leading order, or $n_s=0.968$--$0.974$ and $r=0.011$--$0.008$, with exact slow roll, for $N_*=50$--$60$. This agrees with current data, including the upward shift of $n_s$ driven by ACT, SPT and DESI. The relation $r\simeq2\sqrt3\,[\tfrac23(1-n_s)]^{3/2}$ does not depend on reheating. The next BICEP/Keck analysis can find first evidence for $r$ at this level, and SO, BICEP Array will either measure it or rule the model out.

\begin{acknowledgments}
This work was supported by the Estonian Research Council via the grants TARISTU24-TK10, TARISTU24-TK3, and the CoE program TK202 ``Foundations of the Universe''.
\end{acknowledgments}

\bibliography{refs}

%======================================================================
\section*{End Matter}

The distortion in terms of torsion and non-metricity is
\begin{multline}\label{eq:distortion}
N_{\lambda\mu\nu}=\tfrac12\big(T_{\lambda\mu\nu}+T_{\nu\lambda\mu}+T_{\mu\lambda\nu}\big)\\
  +\tfrac12\big(Q_{\lambda\mu\nu}-Q_{\nu\lambda\mu}-Q_{\mu\lambda\nu}\big),
\end{multline}
with the first index lowered by $g_{\mu\nu}$. With $W_\mu\equiv\Qt_\mu-\tfrac14Q_\mu$, the projectively invariant tensors are
\begin{align}\label{eq:traceless}
  \mathcal T_{\lambda\mu\nu}&=T_{\lambda\mu\nu}+\tfrac13\big(g_{\lambda\nu}T_\mu-g_{\lambda\mu}T_\nu\big),\nonumber\\
  \mathcal Q_{\lambda\mu\nu}&=Q_{\lambda\mu\nu}-\tfrac14g_{\mu\nu}Q_\lambda .
\end{align}
Removing their axial and vector parts,
\begin{align}\label{eq:hats}
  \hat T_{\lambda\mu\nu}&=\mathcal T_{\lambda\mu\nu}-\tfrac16\,\epsilon_{\lambda\mu\nu\rho}\Tt^\rho ,\\\label{eq:hats_2}
  \Qh_{\lambda\mu\nu}&=\mathcal Q_{\lambda\mu\nu}-\tfrac29\big(g_{\lambda\mu}W_\nu+g_{\lambda\nu}W_\mu-\tfrac12g_{\mu\nu}W_\lambda\big).
\end{align}
The tensor $\hat T_{\lambda\mu\nu}$ is traceless with $\hat T_{[\lambda\mu\nu]}=0$, and $\Qh_{\lambda\mu\nu}$ is totally traceless. The quadratic form in Eq.~\eqref{eq:Rdec} is
\begin{equation}\label{eq:quadratic}
  \hat{R}=\tfrac12\hat T_{\lambda\mu\nu}\hat T^{\lambda\mu\nu}+\tfrac14\Qh_{\lambda\mu\nu}\Qh^{\lambda\mu\nu}
  -\tfrac12\Qh_{\lambda\mu\nu}\Qh^{\mu\lambda\nu}+\Qh_{\lambda\mu\nu}\hat T^{\mu\lambda\nu}.
\end{equation}
On the 48 tensor components $\hat R$ is non-degenerate, with 22 positive and 26 negative directions, so the tensor parts of the distortion vanish whenever they have no source.

\end{document}